\documentclass[submission,copyright,creativecommons]{eptcs}
\providecommand{\event}{MeTRiD 2018} % Name of the event you are submitting to
\pdfoutput=1
\usepackage{underscore}           % Only needed if you use pdflatex.

\usepackage{graphicx}
\usepackage{amsmath}
\usepackage{amssymb}
\usepackage{multirow}

\title{Formal Verification of Usage Control Models: \\A Case Study of UseCON Using TLA+} 
\author{Antonios Gouglidis
\institute{School of Computing and Communications\\
Lancaster University\\
Lancaster, United Kingdom
}
\email{a.gouglidis@lancaster.ac.uk
}
\and
Christos Grompanopoulos 
\institute{Department of Mechanical Engineering\\
University of Western Macedonia\\
Kozani, Greece
}
\email{cgrompanopoulos@uowm.gr
}
\and
Anastasia Mavridou 
\institute{Institute for Software Integrated Systems\\
Vanderbilt University\\
Nashville, TN, USA
}
\email{anastasia.mavridou@vanderbilt.edu
}
}
\def\titlerunning{Formal Verification of Usage Control Models}
\def\authorrunning{A. Gouglidis, C. Grompanopoulos, \& A. Mavridou
}
\begin{document}
\setlength{\marginparwidth}{2.2cm}

\maketitle

\begin{abstract}
Usage control models provide an integration of access control, digital rights, and trust management. To achieve this integration, usage control models support additional concepts such as attribute mutability and continuity of decision. However, these concepts may introduce an additional level of complexity to the underlying model, rendering its definition a cumbersome and prone to errors process. Applying a formal verification technique allows for a rigorous analysis of the interactions amongst the components, and thus for formal guarantees in respect of the correctness of a model. In this paper, we elaborate on a case study, where we express the high-level functional model of the UseCON usage control model in the TLA+ formal specification language, and verify its correctness for $\leq12$ uses in both of its supporting authorisation models.
\end{abstract}

\section{Introduction}
Access control systems offer the mechanisms to control and limit the actions or operations that are performed by a user or process -- referred to as \emph{subjects} -- on a set of system \emph{objects}. Specifically, an authorisation process is required to take a decision for granting or denying a \emph{subject} to access an \emph{object} of the system, based on a set of existing policy rules. Thus, access control systems are considered to be amongst the most critical of security components. Their importance is recently highlighted in NIST's special publication \cite{hu2017verification}, where verification approaches for various access control models are examined. 

New computing paradigms, e.g., the cloud \cite{gouglidis2014security}, require further investigation of access control systems. This eventually introduces the need for models, such as usage control \cite{grompanopoulos2012use}, to cope with complex high-level requirements set by new computing paradigms.Usage control models provide an integration of access control, digital rights, and trust management, which may be applicable in environments such as the cloud. To achieve this integration, usage control models support additional concepts such as attribute mutability and continuity of decision. Attribute mutability is responsible for updating the values (e.g., attribute values) of the participating entities (e.g., \emph{subjects}, \emph{objects}), which results in having a dynamic -- usage -- control model. Continuity of decision considers that access to an \emph{object} is no longer an instantaneous access or action, but it may last for some time. Consequently, decision factors are evaluated not only before (i.e., pre-authorisation), but also during the exercise of an access (i.e., ongoing-authorisation), and thus evolving the concept of access control to that of usage control.

Temporal logic is applied to provide unambiguous semantics to functions supported by usage models, such as attribute mutability and continuity of decision \cite{lazouski2010usage}. Usage control systems are concurrent and characterised by non-determinism due to the potential of a \emph{subject} to arbitrarily request or terminate using an \emph{object} during the operation of a system. As discussed in \cite{zhang2008toward}, a number of usage control formal models are limited to the specification of a single use, which is isolated and has no interference with other uses. UseCON \cite{grompanopoulos2012use} is a usage model that provides enhanced expressiveness compared to existing access/usage control models and applicable in new computing paradigms \cite{grompanopoulos2012use}. This is achieved by introducing a new entity entitled \emph{use}, which enables the use of historical information in usage control decision. To ensure the correctness of the UseCON model, we used an existing formal language to verify the correctness of its supported \emph{use} management procedures. For example, ensure out of bound values are not assigned to attribute values, and that authorisation states of the usage model always comply with certain behaviours and follow predefined authorisation transitions (e.g., an authorisation must always be requested before changing into another state). The enhanced version of Temporal Logic of Actions (TLA+)~\cite{Lamport2002} is selected for the formal specification of UseCON -- TLA+ has been selected also in formal definitions of other usage models (e.g., UCON \cite{Zhang2004}). In this paper, we anticipate that the formal specification and verification of UseCON using an automated and error-free model checking technique will provide a comprehensive and unambiguous understanding of the concepts introduced in UseCON and ensure its correctness for a number of uses, depending on the authorisation model (pre-authorisation and ongoing-authorisation). 

The remainder of this paper is organised as follows: Section 2 provides a specification of UseCON's high-level requirements in TLA+, and the verification of its system model is described in Section 3. A performance evaluation of the TLC model checker to verify deadlock, safety and liveness properties is given in Section 4. Concluding remarks and future work are provided in Section 5. An overview of UseCON and TLA+ specifications are available in \cite{grompanopoulos2012use}.    

\section{Specification of UseCON in TLA+}
\label{usecon}
UseCON is a usage control model that is characterised by extended expressiveness when compared with existing usage-based models (e.g., UCON \cite{Zhang2004}). Specifically, it may support authorisations that have logical relations between entities (see \textit{direct} and \textit{indirect} entities in  \cite{grompanopoulos2012use}), as well as historical information, and can express complicated requests between \textit{subjects} and \textit{objects} through simple policies (see automated management of use entities in \cite{grompanopoulos2012use}). In the following, we provide a formal specification of the UseCON model in TLA+ (see TLA+ specification in \cite{TLATools}). This includes the specification of UseCON's main elements, decision making rules, and procedures for managing \emph{use} elements. Furthermore, we describe two transition systems ($TS$) of UseCON, i.e., one $TS$ for the \emph{pre} and one $TS$ for the \emph{ongoing} authorisation model, which are used for supporting concurrent operations amongst uses in UseCON. 

\subsection{Main Elements} 
A \textit{use} in UseCON represents a request to execute an \textit{action (a)} from a \textit{subject (s)} on an \textit{object (o)}. All \textit{subjects}, \textit{objects} and \textit{actions} used in the usage control system define the sets of \textit{subjects (S)}, \textit{objects (O)}, and \textit{actions (A)}, respectively.  The set of \textit{subjects} $S$, \textit{objects} $O$, and \textit{actions} $A$ are defined, as follows: 

\begin{equation*}
S \triangleq \lbrace s_{1}, \ldots, s_{i}\rbrace,  \\
O \triangleq \lbrace o_{1}, \ldots, o_{j}\rbrace,  \\
A \triangleq \lbrace a_{1}, \ldots, a_{k}\rbrace \\
\end{equation*} 

The set of entities E is defined as the union of $S$, $O$ and $A$ sets, as follows:
\begin{equation*}
E \triangleq S \cup O \cup A
\end{equation*} 

The characteristics of an entity are represented through its attribute values. An attribute is a function whose domain is a particular set ($S$ or $O$ or $A$) and its range is composed of specific attribute values, as follows:

\begin{equation*}
att_i:X \mapsto RangeAtt_i
\end{equation*}

\noindent where $X = S$, or $X = O$, or $X = A$.
For every system's entity, an identification attribute \emph{id} is defined for assigning a unique value to the entity. The \emph{id} value remains the same throughout the life-time of the usage control system. Thus, the following invariant is valid for all the behaviours of the usage control system:

\begin{equation*}
\forall \: e_{1}, e_{2} \in E \colon id[e_{1}]=id[e_{2}] \Rightarrow e_{1}=e_{2}
\end{equation*} 
 
In UseCON, the operation of the usage control system does not modify automatically the attribute values of system entities. Thus, attribute values of system entities are proposed to be updated manually by the usage control administration model. Consequently, a system entity is represented with a constant record having as fields the entity's attribute values, as follows:

\begin{equation*}
e \triangleq [ id(e) = k, att_{1}(e) = l_{1}, att_{2}(e) = l_{2}, \ldots , att_{n}(e) = l_{n}]
\end{equation*}

\noindent where $e \in E$ is a system entity and $att_{i}(e)$, with $i \in 1,\ldots,n $, represents the value $l_i$ of one of its attributes. The first record field of each entity is an \emph{id} attribute.

A \textit{use} request in UseCON results into the creation of a \emph{use}. A \emph{use} is an entity instantiated by UseCON and describes all the usage requests in a system and it is described with attributes. More specifically, every \emph{use} must contain a \emph{use} id attribute having a tuple composed of the attribute values \emph{sid}, \emph{oid}, \emph{aid} that describe the identities of the \textit{subject}, \textit{object}, and \textit{action} participating in the instantiated \textit{use}. A special attribute \emph{st} is associated to every \emph{use} instance representing the status of the \textit{use}. Its value may be one of the following: \emph{'requested'}, \emph{'activated'}, \emph{'denied'}, \emph{'stopped'}, and \emph{'completed'}. 

A \emph{use} \emph{u} that instantiates a specific \textit{use} request from subject \emph{s} to object \emph{o} for action \emph{a} is represented in the specification of the model with a variable record, having as fields its attribute values as follows: 

\begin{equation*}
 u \triangleq [ sid(u) = s.id, oid(u) = o.id, aid(u) = a.id, st(u) = state, \\ 
 att_{1}(u) = v_{1}, att_{2}(u) = v_{2}, \ldots , att_{n}(u) = v_{n}]
\end{equation*}

\noindent where s.id, o.id and a.id are the identity values of the subject, the object, and the action, respectively. The state attribute \emph{st} gets a value \emph{state} which belongs to the following set: $ state \in \lbrace$ 'requested', 'activated', 'denied', 'completed', 'stopped' $\rbrace $ and $att_{i}(u), i=1,2,\ldots,n$ are \textit{use} attribute values. The set of all system \textit{uses} is $U$. During the operation of the usage control system, $U$ is populated due to the \textit{use} requests. Moreover, as these \textit{use} requests are served by the usage control system, the \textit{use} attribute values are modified. Specifically, the $U$ is altered when a new \textit{use} is requested or change its progress status (e.g., from \textit{'activated'} to \textit{'stopped'}). Consequently, the only variable utilised in the specification is $U$, which is the set containing all the \textit{uses} operated in the UseCON model and it is declared as follows:

\begin{equation*}
VARIABLES \:\:\:\: U
\end{equation*}

In the beginning of a TLA+ specification, several modules may be included for supporting different operators. Our specification includes \emph{Integers} and \emph{FiniteSets} modules, which encompass arithmetic and set-related operators, like \emph{Cardinality}. This is declared as follows:

\begin{equation*}
EXTENDS \:\:\:\: Integers, FiniteSets \\
\end{equation*}

\subsection{Decision Making in UseCON}
Policy rules in UseCON provide an enhanced utilisation of information from entity and \textit{use} attribute values. More precisely, the general form of a UseCON policy rule that governs the allowance of a \textit{use} request from a \textit{subject} $s$ on an \textit{object} $o$ with an \textit{action} $a$, is a boolean valued expression defined as follows: 

\begin{equation*}
Policy\_Rule(s,o,a,S,O,A) \triangleq expression(e_{1},\dots ,e_{n})
\end{equation*}

\noindent where $s, o, a$ are the particular \textit{direct} entities of the \textit{use}. In addition, two or more UseCON policy rules can be combined together with logical operators as follows:
\begin{equation*}
p = p_1 \otimes p_2 \otimes \ldots \otimes p_n 
\end{equation*}
\noindent where $\otimes$ is a logical operator (e.g., AND, OR), and $p_i$ is a policy rule, where $i=1,\dots,n$.

The parameters $e_{i} \colon i \in 1,\dots,n$  of a policy rule that are utilised for the evaluation of the \emph{expression} may have various origins, and thus lead to the creation of the following categories of policy rules:

\begin{itemize}
\item \emph{Direct Policy Rules}: The parameters $e_{i}$ in the expression of a direct policy rule are values only from attributes of \emph{direct} entities or constant values. Specifically, all parameters $e_{i}$, are defined by the following formula:

\begin{equation*}
e_{i} \in \lbrace s,o,a \rbrace \ or \ e_{i} \triangleq l
\end{equation*} 

\noindent where $l$ is a constant value.

\item \emph{Indirect Policy Rules}: The expression of an \textit{indirect} policy rule consists of attribute values stemming not only from \textit{direct}, but also from \textit{indirect} entities. In this case a logical relation exists between the two types of entities, which can be retrieved by a \emph{select} expression. The definition of such an expression is as follows:

\begin{equation*}
e_{i} \triangleq CHOOSE \: x \: \in \: E \colon select(x,s,o,a,l) 
\end{equation*} 

An example of an indirect policy rule could be an expression that evaluates into true or false, if the father of a child has a premium membership. In this example, the child is a \emph{direct} entity, while the father is an \emph{indirect}. The \emph{select} expression should take into account the fact that there is a \textit{'father'} attribute in the child entity having the attribute value of her father's identity.

\item \emph{Complex Indirect Policy Rules} New computing paradigms introduce complex access control policies, where the usage decision is based on information related not only to a single entity, but with a subset of entities. Such complex policies can be supported in UseCON through complex indirect policy rules. More specifically, a parameter $e_{i}$ of a complex indirect policy rule can be, apart from a single (i.e., direct or indirect) attribute value, an aggregation of information. This information is derived from all the entities that satisfy a desired ($select$) predicate. The semantics of a parameter $e_{i}$ of a complex indirect policy rule is as follows:

\begin{equation*}
e_{i} \triangleq aggregation(\lbrace e \in E \colon select(e) \rbrace)
\end{equation*} 

An example of a complex indirect policy rule is one that confirms that the balance sum of all the accounts of a bank customer is over a specific amount. Information that is related with a set of bank accounts, those belonging to the corresponding user, is required for the evaluation of the aforementioned policy rule. Consequently, the \emph{selection} expression defines the subset of the bank accounts that belongs to the specific customer. 
\end{itemize}

\subsection{Use Attribute Update Procedures}
\label{subsec:attupdate}
Attribute values are utilised by UseCON policy rules through the usage decision making process. Consequently, the implementation of a high level policy should also cope with the definition of \textit{use} attribute value update procedures because these values determine the outcome of the policy rule. In UseCON the mutation of attribute values only records security related information that are related with the \textit{use}. Moreover, a categorisation of the use attributes according to the nature of information they record is as follows:

\begin{itemize}
\item  \emph{Induced Attributes}: Information that can be inferred from entities involved in a \textit{use} (i.e., \textit{subject}, \textit{object}, and \textit{action}) that a specific use instantiates. An example of an \emph{induced} use attribute could be the price of an offered service. 
\item \emph{Observed Attributes}: Information recorded during the exercise of a \textit{use}. Such info cannot be derived directly from the entities involved in the \textit{use}. Examples of \emph{observed} use attributes are the duration of a \textit{use}, etc.
\end{itemize}

Update procedures for \emph{induced} attributes can be specified during the definition of the implementation of a high level policy by the policy administrator. However, update procedures for \emph{observed} attributes require the existence of a system function that returns the required value. For example, the specification of a policy, which requires recording the system time whenever a \textit{use} is permitted, is as follows:

\begin{equation*}
preUpdate \triangleq [ u \ EXCEPT \ !.st = ``activated", \ !.allowedtime=SystemTime()]
\end{equation*}

\noindent where \emph{SystemTime()} is an internal function provided by the system framework that provides the current time. In TLA+, \emph{EXCEPT} is a special purpose operator representing the modification of a function from a state to the next. In that next state all function values are left unchanged unless stated otherwise~\footnote{For a comprehensive definition of EXCEPT operator we refer the reader to \cite{Lamport2002}.}.

\begin{figure}
\centering
\includegraphics[scale=0.7]{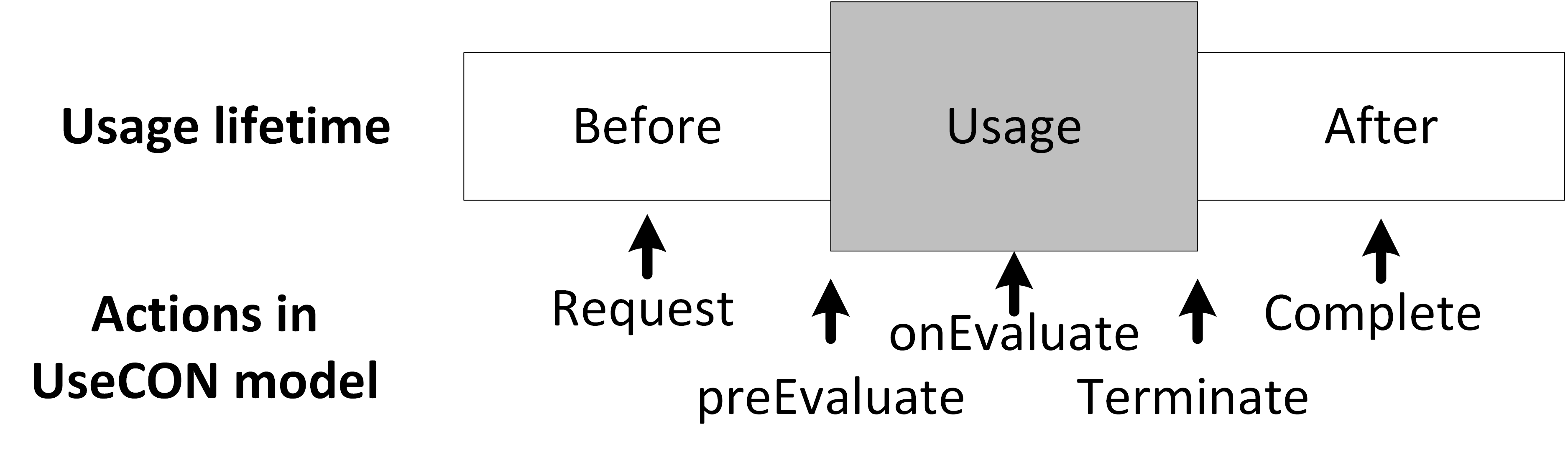}
\caption{Use attributes updates during the exercise of a \textit{use} in UseCON}
\label{fig:updates}
\end{figure}

Attribute update procedures in UseCON are performed whenever a \textit{use} changes its state (e.g., from \textit{`requested'} to \textit{`activated'}). Therefore, during the execution of a \textit{use}, the times of use attribute update in UseCON are represented in Figure \ref{fig:updates}.

\subsection{Transition Systems}
Actions in the UseCON model are categorised to those triggered by a \textit{subject's} request and those operated automatically by the usage control system. More specifically, for every \textit{use} supervised by the usage control system the following actions can be triggered by a \textit{subject}. 

\begin{itemize}
\item \emph{Request}: This action performs the transition from the \textit{`init'} state of the \textit{use} to \textit{`requested'}. Moreover, \emph{request} creates a particular use instance that instantiates the requested \textit{use} and also assigns values to the \textit{id} attribute of the \textit{use}.   

\item \emph{Complete}: This action changes the state of the \textit{use} from \textit{'activated'} to \textit{'completed'}.
\end{itemize}

The actions performed automatically by the usage control system, follows:
\begin{itemize}

\item  \emph{preEvaluate}: This action is performed by the usage control system only when the allowance of the \textit{use} is governed by a pre-authorisation rule. This action changes the state of the \textit{use} to either \textit{'activated'} or \textit{'denied'}, depending on the outcome of the examined policy rule. 

\item  \emph{onEvaluate}: In case the allowance of a \textit{use} is governed by an ongoing authorisation rule, the \emph{onEvaluate} action is performed by the usage control system. If the particular policy rule is satisfied then the state of the \textit{use} does not change. In case the policy rule is not satisfied, the state of the \textit{use} is changed to \textit{'stopped'}. 

\item  \emph{Activate}: This action is performed only when the allowance of the \textit{use} is governed by an ongoing authorization rule. It follows the execution of the \emph{Request} action and changes the state of the \textit{use} from \textit{`requested'} to \textit{`activated'}.
\end{itemize}

\begin{figure}
\centering
\includegraphics[scale=0.85]{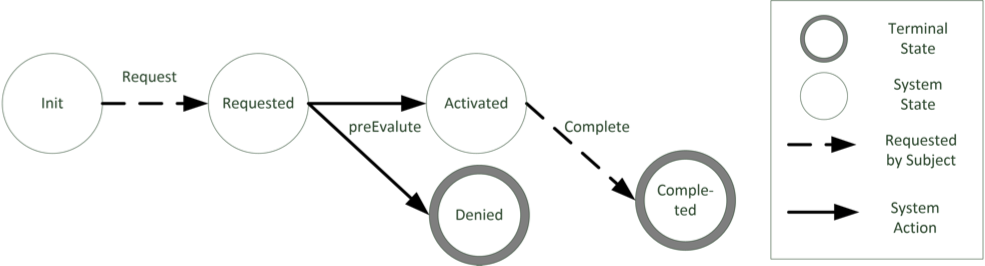}
\caption{Transition system of a single \textit{use} pre-authorisation UseCON model}
\label{fig:preSuTS}
\end{figure}

\begin{figure}
\centering
\includegraphics[scale=.85]{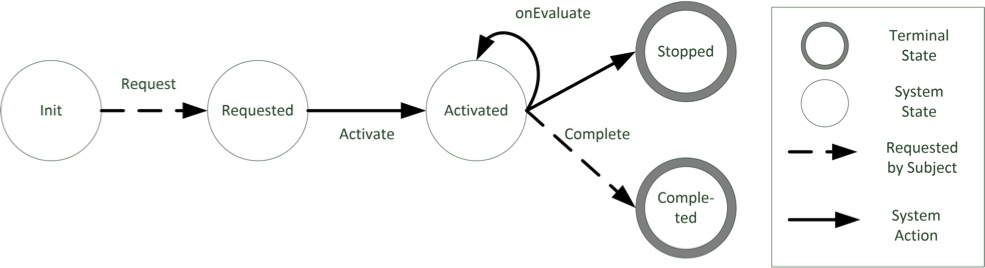}
\caption{Transition system of a single \textit{use} ongoing authorisation UseCON model}
\label{fig:onSuTS}
\end{figure}

Any of the previous actions, apart from modifying the \emph{st} attribute value, may also update other \textit{use} attribute values. Such modifications in the use attribute values are considered to be implementation specific, as already described in Section \ref{subsec:attupdate}. Moreover, despite the existence of a great number of updates on attribute values, the execution of any of the previous actions is considered to be atomic~\footnote{This constraint is realistic due to the fact that a use is recorded centrally on the policy decision point and there is no need for attribute updates of other entities (\textit{subject} or \textit{object}).}, i.e., a single behavioural step. The transition system for a single \textit{use} UseCON system controlled by a pre and an ongoing authorisation policy rule is depicted in Figures \ref{fig:preSuTS} and \ref{fig:onSuTS}, respectively.

In its initial state, no \textit{uses} are exercised in the system, and thus the first state of every behaviour must satisfy the TLA+ predicate \emph{init}, which defines that the set-variable $U$ is an empty set:

\begin{equation*}
Init \triangleq U = \lbrace \: \rbrace
\end{equation*}

Moreover, according to the time period that a \textit{use} request evaluation is performed, a pre-authorisation and ongoing-authorisation transition systems is created.

\subsubsection{Pre-authorisation}
The possible actions that can be performed on a pre-authorisation UseCON system are the \textit{use} request for a new \textit{use}, the evaluation of an already requested \textit{use}, or the termination of a \textit{use} that is already executed. Therefore, the \emph{next} action, that describes all the possible next states could be a \emph{request}, \emph{evaluate}, or \emph{complete} action, described as follows:
\begin{equation*}
Next \triangleq \: Request \vee preEvaluate \vee Complete
\end{equation*}
More specifically, the \emph{Request} action selects non-deterministically~\footnote{The non-determinism property is implied by the use of $\exists$ operator. For comprehensive information refer to \cite{Lamport2002}.} a new \textit{use} $x$. This is verified by searching the set of uses $U$, and returning one that has not already been requested by the subjects or processed by the usage control system. Consequently, in case that this particular $x$ exists, the \emph{request} action creates the corresponding \textit{use} instance that instantiates \textit{use} $x$ and inserts it to the set $U$. The definition of the \emph{Request} action is as follows:
\begin{align*}
Request \triangleq \: \exists & \: u \in (S \times A \times O) \colon ( \\
& \: \: \wedge \: \forall \: x \in U \colon (x.sid \neq u[1].id \vee x.aid \neq u[2].id \vee x.oid\neq u[3].id) \\
& \: \: \wedge U' = U \cup \lbrace createUse(u)\rbrace )  
\end{align*}
The \emph{preEvaluate} action examines if there are any \textit{uses} that have been requested but have not been processed by the usage control system. More specifically, \emph{preEvaluate} examines if there is any use instance with state attribute value equals to \textit{`requested'}. Consequently, the action evaluates the policy rule that governs the allowance of the \textit{use} that the specific use instance instantiates. Based on the outcome of that policy rule the action modifies the state of the use either to \textit{`activated'} or to \textit{`denied'} with \emph{preUpdate} and \emph{denUpdate} use attribute update procedures respectively as follows:  
\begin{align*}
preEvaluate  \triangleq \: \exists \: u \in U \colon (  &\wedge \: u.state=``requested"\\
& \wedge IF(PolicyRule) \: THEN \\
& \qquad U' = ( U \setminus \lbrace u \rbrace ) \cup \lbrace preUpdate(u) \rbrace  \\
& \quad ELSE \\
& \qquad U' = ( U \setminus \lbrace u \rbrace ) \cup \lbrace denUpdate(u) \rbrace ) \\
\end{align*}

The \emph{Complete} action simulates a \textit{subject's} request to terminate the execution of a currently active \textit{use}. If such a use exists, its state attribute value should be equal to \textit{'activated'}. Consequently, the \emph{completed} action modifies the state attribute value from \textit{'activated'} to \textit{'completed'} with the \emph{comUpdate} use attribute update procedure. The \emph{Complete} action is defined as follows: %\Anastasia{Instead of the word semantics throughout the paper, I think we should use defined.}:

\begin{align*}
Complete \triangleq \: \exists \: u \in U \colon ( & \wedge \: u.state=``activated" \\
& \wedge U' = ( U \setminus \lbrace u \rbrace ) \cup \lbrace comUpdate(u) \rbrace ) 
\end{align*}

\emph{CreateUse} is the procedure that creates a use instance that has the identities of the \emph{direct} \textit{subject}, \textit{object}, \textit{action}. The definition of the \emph{createUse} procedure follows:

\begin{align*}
createUse(x) \triangleq [ & sid(x) = x[1].id, aid(x) = x[2].id, oid(x) = x[3].id, \\
& state(x) = requested, att(x) = k]
\end{align*}

All the other use attribute update procedures perform a dual role. Firstly, they alter the \emph{state} use attribute to the desired value (e.g., \emph{preUpdate} to \textit{`activated'} or \emph{comUpdate} to \textit{`completed'}). Secondly, the use attribute update procedures modify the values of other use attributes according to the requirements of the usage control system that they are called to describe. For example, a possible \emph{preUpdate} procedure can be the following:
\begin{equation*}
preUpdate \triangleq [ u \ EXCEPT \ !.st=``activated", \ !.att=value]
\end{equation*}

The same definitions apply to the \emph{denUpdate, comUpdate} procedures.

\subsubsection{Ongoing-authorisation}
The transition system of the ongoing-authorisation UseCON model is differentiated from the pre-au\-tho\-ri\-sa\-tion model in a number of ways. Firstly, in an ongoing model, a \textit{use} that is requested is permitted to be activated without the evaluation of any policy rule. Secondly, at a given time interval~\footnote{The determination of the exact interval is left open as an implementation issue.}, an ongoing action is executed \emph{onEvaluate}. Thus, the specification of the \emph{Next} action on an ongoing authorisation model has the following definitions: 
\begin{align*}
Next \triangleq \: Request \vee Activate \vee onEvaluate \vee Complete
\end{align*}

The \emph{Activate} action searches for the existence of a use with state attribute value equal to \emph{'requested'} and consequently updates it to \emph{`activated'} by executing the \emph{preUpdate} procedure.
\begin{align*}
Activate \triangleq \: \exists \: u \in U \colon ( & \wedge \: u.state=``requested" \\
& \wedge U' = (\lbrace U \setminus \lbrace u \rbrace ) \cup  \lbrace preUpdate(u) \rbrace )  
\end{align*}

Whereas \emph{onEvaluate} action evaluates an ongoing policy rule, based on this result it either leaves use to \textit{`activate'} state, but it can possibly update the rest use attribute values with \emph{onUpdate} procedure, or modify its state attribute value to \textit{`stopped'} with \emph{termUpdate} use attribute update procedure as follows: 
\begin{align*}
onEvaluate \triangleq \: \exists \: u \in U \colon ( & \wedge \: u.state=``activated" \\
& \wedge IF(PolicyRule) \: THEN \\
& \qquad U' = (\lbrace U \setminus \lbrace u \rbrace ) \cup  \lbrace onUpdate(u) \rbrace  \\
& \quad ELSE \: \\
& \qquad U' = (\lbrace U \setminus \lbrace u \rbrace ) \cup  \lbrace stopUpdate(u) \rbrace ) 
\end{align*}
Moreover, in the case where there is no need for an ongoing use attribute update procedure, the \emph{onUpdate} procedure can be substituted with the \emph{UNCHANGED U} TLA+ operator that leaves the variable $U$ unmodified. Here, the semantics of \emph{`request'} and \emph{`complete'} actions are considered to be the same with the pre-authorisation model.

\section{Model Checking with TLC}
\emph{Toolbox} is an Integrated Development Environment (IDE), which is designed for the definition and verification of TLA+ specifications \cite{TLATools}. Specifically, the \emph{toolbox} editor provides functionality for the definition and alteration of TLA+ specifications, and supports syntax highlighting. Additionally, an automatic parser checks the defined specifications for syntax errors and presents them accordingly by marking them in the used modules.

The tool in use for the verification of a TLA+ specification in \emph{toolbox} is the TLC model checker. Specifically, TLC explicitly generates and computes all the possible states of a system. However, many times the specification of a system might contain an infinite number of states. TLC handles such specifications, by choosing a finite model of the system and in turn checks it thoroughly. Specifically, the creation of a system's model in TLC requires the definition of its specifications, properties and values of constant parameters \cite{TLATools}. A specification represents all the behaviours that have to be checked. Moreover, the values assigned to constant parameters are utilized for the instantiation of a specification. TLC can check a model for deadlocks, invariants and properties \cite{TLATools}. A deadlock occurs when the model reaches a state in which its next-state action allows no successor states. An invariant is a predicate that is evaluated on a system state. Consequently, an invariant holds on a system specification, if and only if, every state of all the behaviours of a system satisfy that predicate. Properties are temporal formulas that must be evaluated to true for all the behaviours of the model. TLC has some limitations regarding the handling of a subclass of TLA+ specifications and properties that it can check \cite{Lamport2002}. A very helpful feature of TLC is the fact that when it identifies an error during the verification process, it provides an error trace viewer that allows the exploration in a structured view of the debugging information. Moreover, TLC supports an arbitrary evaluation of states and action formulas in each step of the trace.

\begin{table}[]
\centering
\caption{Safety and liveness properties in UseCON}
\label{tbl:VerifTable}
\begin{tabular}{cllll}
\hline
\multicolumn{1}{l}{}              & \multicolumn{2}{c}{\textbf{Safety}}           & \multicolumn{2}{c}{\textbf{Liveness}}         \\ \hline
\multicolumn{1}{l}{}              & \textbf{Former state} & \textbf{Latter state} & \textbf{Former state} & \textbf{Latter state} \\ \hline
\multirow{3}{*}{\textbf{Pre}}     & Completed             & Any other state       & Requested             & Activated or Denied   \\
                                  & Activated             & Requested or Denied   & Requested             & Completed or Denied   \\
                                  & Denied                & Any other state       & Activated             & Completed             \\ \hline
\multirow{3}{*}{\textbf{Ongoing}} & Completed             & Any other state       & Requested             & Activated             \\
                                  & Activated             & Requested             & Requested             & Completed or Stopped  \\
                                  & Stopped               & Any other state       & Activated             & Completed or Stopped \\ \hline
\end{tabular}
\end{table}

\subsection{Use management}
\label{sec:UseMan}
One of the fundamental properties that can be verified in a system is that of type correctness. Specifically, type correctness is considered to be an invariant which  determines that all the variables of the system are assigned with values originating only from a specific set of values. The UseCON specification uses a single variable $U$ which corresponds to the set of system uses. The \emph{invariant} property that defines type correctness in the UseCON model, is defined as follows:
\begin{align*}
& TypeCorrectness \triangleq U \subseteq Uses 
\end{align*}

\noindent where \textit{Uses} is the set of all records that have the following form (i.e., all the record fields are assigned with values originating from their domains):
\begin{align*}
[ s.id \colon SubjectIDS, o.id \colon ObjectIDS, a.id \colon ActionIDS, st \colon USTATE, att \colon ATTDOMAIN ]
 \end{align*}

Moreover, the definition of the domains \textit{SubjectIDS, ObjectIDs, ActionIDs} and \textit{USTATE} is:
\begin{align*}
 SubjectIDs \triangleq \lbrace s.id \colon s \in S \rbrace,
 ObjectIDs \triangleq \lbrace o.id \colon o \in O \rbrace, 
 ActionIDs \triangleq \lbrace a.id \colon a \in A \rbrace 
\end{align*}
\noindent where
\begin{align*}
 USTATE \triangleq \lbrace \text{requested, activated, denied, stopped, completed} \rbrace 
\end{align*}

A \emph{use} is capable of recording detailed historical information about the operation of \textit{uses} in the system. Consequently, a valid implementation of the UseCON model, where multiple \textit{use} processes are operating concurrently, depends on a proper management of the \textit{use} instances that represent these \textit{uses}. Specifically, all \textit{use} instances must adhere only to the state transitions depicted in Figures \ref{fig:preSuTS} and \ref{fig:onSuTS} for the pre-authorisation and ongoing-authorisation models, respectively. Based on that, a number of \emph{safety} and \emph{liveness} properties can be defined. For example, a \emph{safety} property (a faulty state cannot be reached) states that a \textit{use} instance cannot be evaluated as \textit{'requested'} in its \textit{st} attribute, if it has previously been evaluated as \textit{'completed'}. The semantics, expressed in TLA+, which verify the previous property for all the uses of a system are defined by the following temporal formula:

\begin{equation*}
\mathit{Safety} \triangleq \square ( \exists \: u \in U \colon u.st=\text{``completed"} \implies \square ( u.st \neq \text{``requested"}) )
\end{equation*}

Similar \emph{safety} properties can be defined for all the possible prohibited state transitions. The complete set of prohibited state transitions is depicted in Table \ref{tbl:VerifTable} under the header \emph{`safety'}. Moreover, two examples that illustrate the violation of \emph{safety} properties are depicted in Figure \ref{fig:safetyEx}. Specifically, the transition of the \emph{use} instance $u1$ from  \textit{`completed'} into \textit{`requested'} results in a violation of a \emph{safety} property. A similar violation is considered during the transition of the \emph{use} instance $u2$ if the state transits from \textit{'activated'} to \textit{`requested'}. In addition, the definition of \emph{liveness} properties in the UseCON model determine all the valid state transitions regarding any use instance. For example, Figure \ref{fig:preSuTS} presents that a \textit{use} instance that has at any given state an \emph{st} attribute value that is evaluated to \textit{`activated'}, then its attributed value must be eventually evaluated as \textit{'completed'}. All the possible state transitions that are eligible to be performed are depicted in Table \ref{tbl:VerifTable} under the header \emph{'liveness'}.
 This property is defined in TLA+ as follows: 

\begin{figure}
\centering
\includegraphics[scale=1]{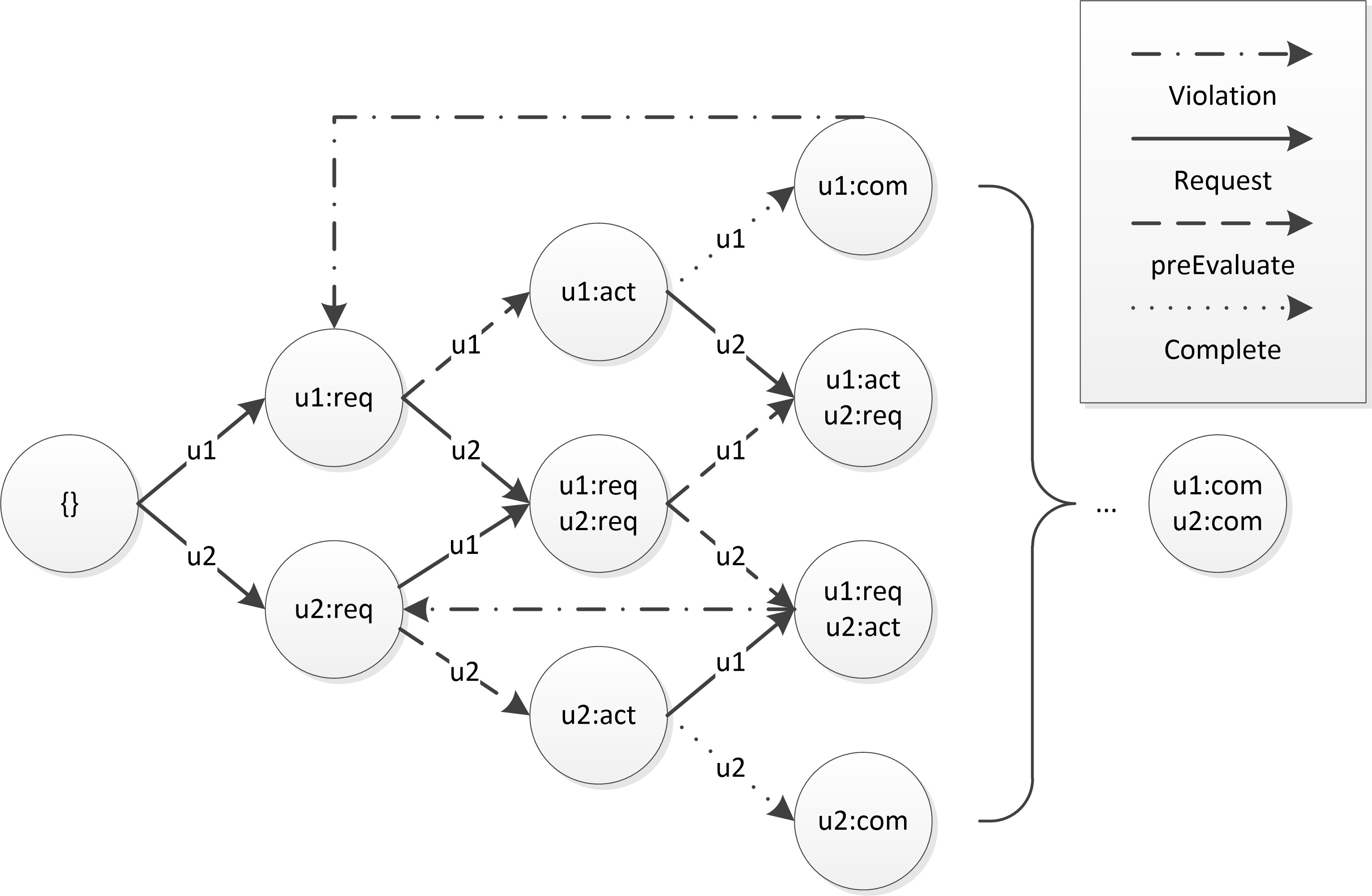}
\caption{Violation of safety properties }
\label{fig:safetyEx}
\end{figure}

\begin{equation*}
\mathit{Liveness} \triangleq \forall \: u \in U \colon u.st=\text{``activated"} \rightsquigarrow u.st = \text{``completed"}\\
\end{equation*}

\section{Performance Evaluation}
\label{subsec:verifresults}
The verification of the examined model was performed on a MacBook Pro (Mid 2014) using TLA+ Tool version 1.5.6 and running TLC version 2.12. The system operates on OS X El Capitan and its hardware specifications are: 2.5GHz Intel Core i7 with 16GB 1600 MHz DDR3.

A first set of results were collected by verifying deadlock-freedom for the model of UseCON. Specifically, in a pre-authorisation UseCON model, we consider all the \textit{uses} of the model to be requested, and therefore, to be in either the state of \textit{`activated'} or \textit{`denied'}. The \textit{uses} being activated were finally completed. Therefore, the final state in every \textit{use} must be \textit{`denied'} or \textit{`completed'}. The TLC model checker evaluates all the behaviours of the model and terminates when it reaches to a deadlock, and  the actions of the deadlocked behaviour are presented along with the attribute values in each state. Moreover, we performed a verification of the model against safety and liveness properties described in Section \ref{sec:UseMan}. The verification was finished without raising any errors, but this applies only for the verification of $\leq12$ uses for both the pre and ongoing authorisation models. Although higher numbers of \textit{uses} were considered, the verification process had to be interrupted due to the excessive amount of time required for its completion given the number of \textit{uses}. The collected verification results (see Table \ref{tbl:times}) requires a better understanding of the internal procedures TLC is applying to compute the behaviours of the model (i.e., generation of the transition system). Initially, TLC computes the states that verify the \emph{Init} predicate and inserts them into a set $G$. For every state $s \in G$, TLC computes all the possible states $t$ that $s \mapsto t$ can be a step in a behaviour. Specifically, TLC substitutes the values assigned to variables by state $s$ for the unprimed variables of the \emph{Next} action, and then it computes all the possible assignment of values to the primed variables that makes the \emph{Next} action true. For every state $t$ found by the former procedure it is added to set $G$ if it does not already exists. The previous two actions are repeated until no new states can be added in $G$. Therefore, the verification results produced by TLC, are: \emph{Diameter} expresses the number of states in the longest path of $G$ in which no state appears twice; \emph{States found} expresses the number of examined states; \emph{Distinct States} expresses the number of examined distinct states. The verification results produced by TLC for the \emph{pre} and \emph{ongoing} UseCON models are presented in Table \ref{tbl:times}. An additional column presents the actual running time of the TLC model checker in seconds.

\begin{table}[]
\centering
\caption{Performance evaluation}
\label{tbl:times}
\begin{tabular}{ccccccc}
\hline
         \textbf{Authorisation}                   & \textbf{Uses} & \textbf{Diameter} & \textbf{States} & \textbf{Distinct} & \textbf{Deadlock} & \textbf{Safety and Liveness}\\ 
         \textbf{model}     &  & & \textbf{found} & \textbf{states} & \textbf{(seconds)} & \textbf{ (seconds)}\\                                
                                \hline
\multirow{3}{*}{\textbf{Pre}}   
& 2	& 6	&		21 &		12 &			1 &		3 \\
& 8	& 23 &			277969 & 	16832 &		2 &		5 \\
& 12 &	31 &			45533665 &  560128 &		53 &		138 \\ \hline

\multirow{3}{*}{\textbf{Ongoing}} 
& 2 & 7 & 33 & 16 & 1 & 3 \\ 
& 8 & 25 & 367873 & 23808 &  2 & 7 \\
& 12 & 37 & 79112449 & 1224704 & 118 & 241 \\ \hline 

\end{tabular}
\end{table}

\section{Conclusion and Future Work}
The application of model checking as a technique resulted in formally verifying the use management functions of UseCON. The trace of the deadlock errors verifies that the defined specifications of the system operate correctly. An advantage of using model checking techniques for the verification of usage control models is the provision of formal guarantees with regards to the correctness of the model, without requiring an implementation of it. Nevertheless, known issues of model checking, i.e., state explosion problem, prevented the timely verification of \textit{uses} when these increase in number. This resulted in providing formal guarantees for the correctness of UseCON for $\leq12$ uses for both the pre-authorisation and ongoing-authorisation models, as depicted in Table ~\ref{tbl:times}. 

In the future, we aim to demonstrate the verification of complex policy rules supported by UseCON, and specifically investigate the verification of ongoing policy rules. This is of interest since a \textit{use} request might lead to a policy violation in other concurrent \textit{uses}. Thus, any \textit{use} request should be followed by an evaluation of all policy rules in all \textit{uses} in the system to avoid conflicts and violations. Finally, we are considering frameworks, such as secBIP~\cite{said2014model}, that will allow us to compositionally analyse security properties and generate secure-by-construction systems. Also, we would like to explore guaranteeing security properties by-construction for any number of \textit{uses} in our model through the application of architectures from predefined architecture styles~\cite{mavridou2016architecture} that capture properties of specific access and usage control policies.

\nocite{*}
\bibliographystyle{eptcs}
\bibliography{generic}

\end{document}